\documentclass[pdftex]{class}
\usepackage{graphicx}
\usepackage{amsmath}
\usepackage{amssymb}
\usepackage{txfonts}
\usepackage{xcolor}
\usepackage{supertabular,lscape,epsfig}

\SetPages{1}{12}
\SetVol{76}{2026}

\begin{document}
\begin{Titlepage}
\Title{Revisiting BV \& BW Draconis: \\Detailed Study of Intriguing Multiple System}

\Author{J. J. T~o~k~a~r~e~k, W. D~i~m~i~t~r~o~v, M. P~o~l~i~\'{n}~s~k~a, M. K. K~a~m~i~\'{n}~s~k~a,\\K. K~a~m~i~\'{n}~s~k~i, P. K~o~l~e~\'{n}~c~z~u~k, J. P~i~e~t~r~u~c~h~a, Sz. P~t~a~k and M. W~i~t~c~z~a~k}
{Astronomical Observatory Institute, Faculty of Physics and Astronomy, A. Mickiewicz University, \\ul. S{\l}oneczna 36, 60-286 Pozna\'{n}, Poland}

\footnotesize{April, 2026}
\end{Titlepage}

\Abstract{The BV \& BW Draconis system is a unique coplanar system of two eclipsing contact binaries.
The main objective of this study is to develop a model based on new observations. The only model of the system was obtained 40 years ago. 
New photometric and spectroscopic data have been collected. Brightness measurements were performed using the differential aperture photometry method, while radial velocity measurements were performed using the cross-correlation method. The resulting curves were analyzed using the Wilson-Devinney method. The resulting model is similar to the one obtained previously; however, the masses of the four components are higher, and the differences are outside the margin of error. In the case of BW Dra, the photometric data indicate the presence of a spot on the star's surface.
Based on our spectroscopic and photometric data, the latest measurements of eclipse times have been obtained. 
These results, together with other new measurements, confirm previous findings: both pairs exhibit eclipse time variability. In the case of BV Dra, the parabolic and sinusoidal trends were confirmed and refined. For the second pair, BW Dra, a change in behavior was observed that may be related to a change in period. 
}
{Stars: individual: BV Dra - Stars: individual: BW Dra}

\section{Introduction}
BV \& BW Draconis is a well-known visual binary, listed as ADS 9537 in Aitken (1932). Their variance in luminosity was first announced in Batten and Hardie (1965), and in the same work they suggest that this system could be not only visually but physically bounded. The first accurate spectroscopy study was conducted by Batten and Lu (1986), and the first full orbital solution including photometry data was provided by Kaluzny and Rucinski (1986). The two visual components of the system have similar brightness (8.06 mag for BV Dra and 8.68 mag for BW Dra, in the Gaia G filter), and both are short-period W UMa-type eclipsing variables, with periods respectively around 0.350 and 0.292 days. The BV Dra eclipsing binary consists of stars with masses 1.00 and 0.40 $M_\odot$ and the BW Dra component masses are 0.89 and 0.25 $M_\odot$ (Batten and Lu 1986). Additionally, in Yang \textit{et al.} (2009) it was suggested that sinusoidal changes in the (O-C) curve for BV Dra can be explained by the presence of a third body, which has an orbital period of P3 = 23.8 $\pm$ 0.6 yr. Altogether with our recent work (Tokarek and Dimitrov 2025), this strongly suggests that the BV\&BW Dra system is a quintuple system. This fact, combined with almost 40 years that have passed since the first (and only) detailed study, motivated us to revisit this case. 

\section{Observations}  

\subsection{Radial velocities}
Instruments used to collect spectroscopic data were fibre-fed echelle spectrographs installed on 2.0m Ritchey-Chretien-Coude (RCC) Telescope at Rozhen National Astronomical Observatory (NAO), Bulgaria, called ESpeRo\footnote{Based on data collected with 2-m RCC telescope at Rozhen National Astronomical Observatory.} (Bonev \textit{et al.}~2017) and Pozna\'n Spectroscopic Telescope 1 (PST1) in Bor\'owiec, Poland (Baranowski \textit{et al.}~2009), which has smaller, 0.5 m aperture. Both instruments are using as detectors 2048$\times$2048 Andor back-illuminated CCD cameras (iKon-L and DZ436, respectively). The exposure times were 600 s (since both systems have periods around 7-8 h, we had to balance signal strength and blurring effect from star's orbital motion). 

The data were obtained during a few nights in years 2024 and 2025. Observations at Rozhen were conducted on 31 March and 22 June 2024, and we observed at PST1 on 15, 18, and 19 March, and 2 and 3 April 2025. In total, we obtained 77 spectra of BV Dra (43 from Rozhen and 34 from PST1) and 108 spectra of BW Dra (56 from Rozhen and 52 from PST1). The typical signal-to-noise ratio for BV Dra was in the range 35-45 for the Rozhen spectrograph and 35-40 for PST1. As the BW Dra system is slightly fainter, the SNR in that case was lower, typically 15-20 for the Rozhen instrument and around 10 for PST1. Spectra were reduced and analysed using IRAF (Tody 1986; Tody 1993) to calculate the radial velocities of the systems' components. 

\subsection{Photometry}
\begin{figure}[ht]
\centering
\includegraphics[width=0.8\textwidth]{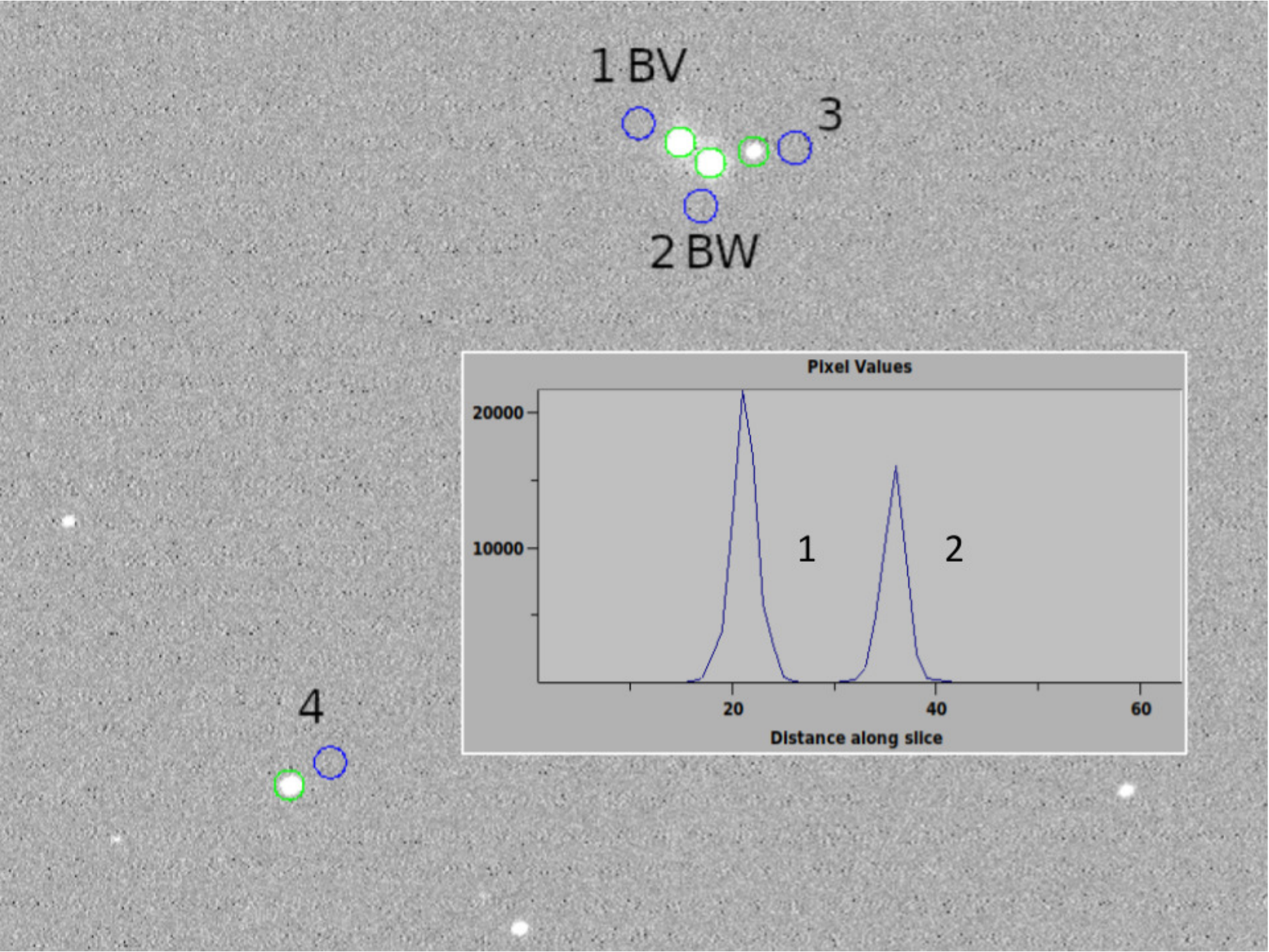}
\caption{The figure shows a part of the image obtained on 20 April 2025 with the RBT telescope. The green apertures labeled 1 and 2 denote BV and BW Dra, respectively, while apertures 3 and 4 indicate the comparison stars. The blue circles indicate the sky apertures used for background determination. The smaller panel presents a cross-section of the BV (stronger peak, 1) and BW (weaker peak, 2) profiles.}
\label{image}
\end{figure}

\begin{figure}[ht]
\centering
\includegraphics[width=0.8\textwidth]{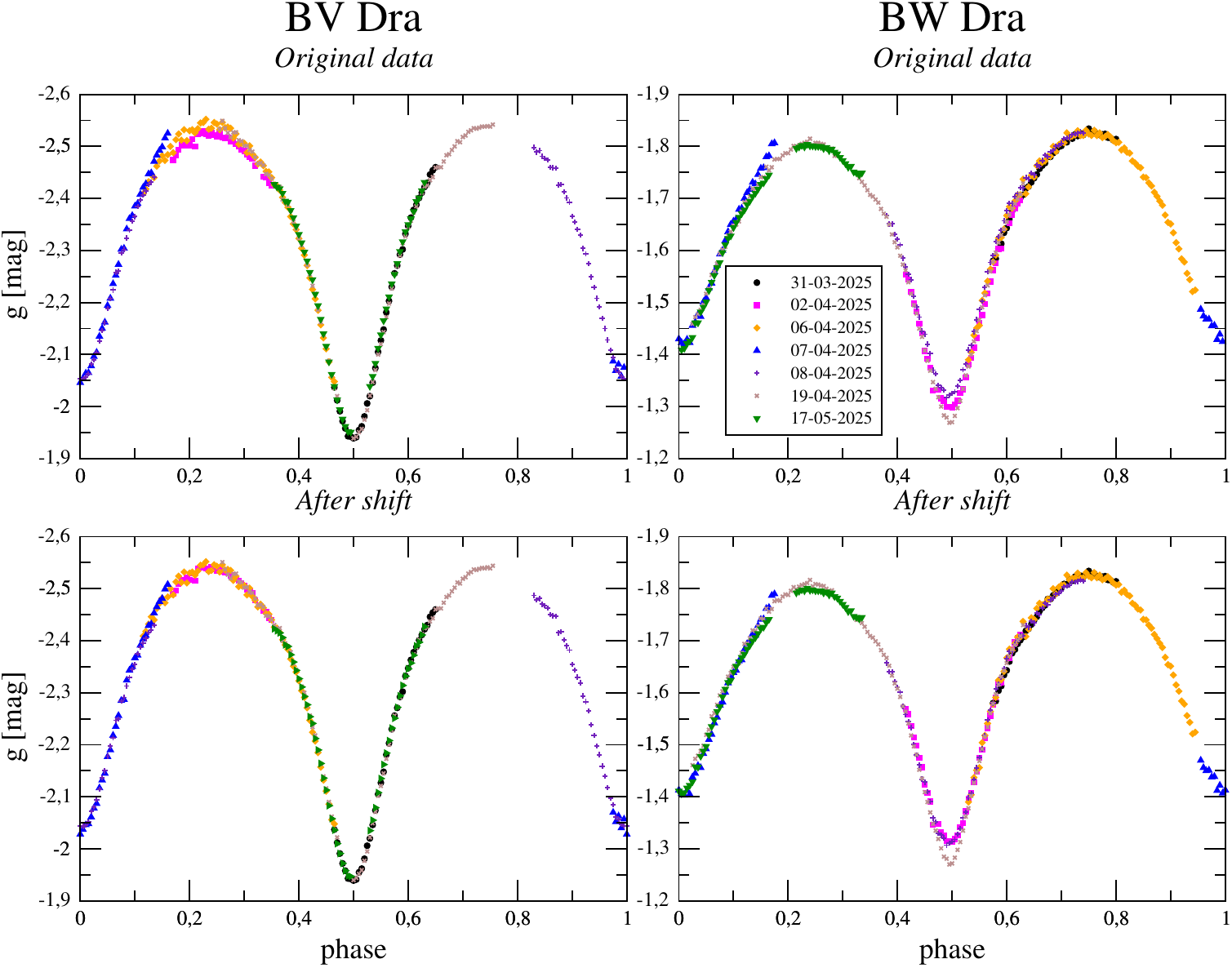}
\caption{Photometric data from RBT before (top row) and after (bottom row) shift. Applied shifts values are in Tab. \ref{shifts_tab}. After shift sum of standard deviations (SD) calculated from each bin changes from 2.3135 to 1.7136, average SD drops from 0.0798 to 0.0527 mag. Periods values used are $p=0.350069$ d for BV Dra, and $p=0.292160$ d for BW Dra}
\label{shifts_pic}
\end{figure}

Due to the small separation on the sky between BV and BW Dra (16 arcsec), we were unable to use photometric data from the TESS mission. The angular resolution of this space mission was insufficient for our needs as the light curves for both stars were mutually contaminated. To complement the spectroscopic observations, we obtained additional photometric observations with the 0.7-m Roman Baranowski Telescope (RBT), located at Winer Observatory, Arizona, USA, and equipped with an Andor iXon EMCCD camera. The telescope is a PlaneWave CDK700 (Corrected Dall-Kirkham) system with an f/6.6 optical design, mounted on an alt-az robotic direct-drive mount with dual Nasmyth foci (Kami\'{n}ski et al. 2014, Poli\'{n}ska et al. 2019). The photometric observations were carried out over seven nights, between 31 March and 17 May 2025 (the exact observing dates are given in the legend box in Fig. \ref{shifts_pic} and \ref{plots}). To avoid saturation and blending of the profiles of the two systems, the PAN-STARRS g filter, which provided the lowest signal level, was used together with the shortest possible exposure time. The exposure time was 1 s on the first night and 0.5 s on each of the remaining six nights. This made it possible to separate the light of the two stars in the photometric observations, as clearly shown in the inset panel of Fig. \ref{image}.

The observations were reduced following the standard procedure using the Starlink Software Collection (Currie et al. 2014), including bias subtraction and flat-field correction. Differential aperture photometry was then performed with GAIA, version Namaka, distributed as part of the Starlink Software Collection. All stars available in the frames were examined in order to exclude stars showing variability. The final relative photometry was performed with respect to the star marked by aperture No. 4, BD+62 1390 (Fig. \ref{image}). This star was chosen because it was the closest to the target stars, one of the brightest stars in the field, and showed no detectable brightness variations. Its spectral type (G0) is also comparable to those of the analysed objects. The same star was also used as a comparison star in the photometry of the BW Dra by Rovithis \& Rovithis-Livaniou (1983). 

The standard sky background annular aperture surrounding the target was not used. Instead, the sky background was determined from separate sky apertures placed close to the stars in source-free regions. This way, the background apertures did not affect light from the closely spaced BV and BW Dra components. The sky background apertures are marked with blue circles in Fig. \ref{image}. The aperture sizes were kept constant throughout each night for all measured stars and ranged from 7 to 9 pixels depending on the night of observation. The width of the stellar profiles is also visible in the cross-section shown in the inset panel of Fig. \ref{image}.

The light curves obtained from photometric observations and phased with the orbital periods separately for BV and BW Dra are shown in Fig. \ref{shifts_pic}.

As we needed to combine observational data from different nights to obtain (mostly) complete light curves for both stars, we need to deal with the problem of proper joining of separate datasets. Weather conditions, small changes in apertures, and unresolved stellar activity can subtly influence the brightness of both BV and BW Dra. We needed to shift the datasets to make them as efficient as possible. To do that, we binned measurements from each observational night in 200 phase intervals for the BV and BW Dra light curves. Next, we calculated the standard deviation (SD) of measurements in each bin, with at least two separate data points (data from different nights), and summed them over both light curves. Then, we applied photometric shifts for each night to minimise this summed up SD. The light curves before and after that process are visible in Fig. \ref{shifts_pic}, and the values of each shift are presented in Tab. \ref{shifts_tab}. This helps during the modelling stage of our study, but does not mitigate all data disturbances, as data for BW Dra from nights 07-04, 19-04, and 17-05 near phase 0.2 are not convergent, and 19-04 data for 0.5 phase (primary minimum) have a visible lower level from the others datasets. 

\MakeTable{l|ccccccc}{12.5cm}{Values of photometric shifts applied to each datasets from separate nights\label{shifts_tab}}
{
 \textbf{Date}        & 31-03-25 & 02-04-25 & 06-04-25 & 07-04-25 & 08-04-25 & 19-04-25 & 17-05-25\\
\noalign{\smallskip}
\hline\noalign{\smallskip}
 \textbf{Shift [mag]} & 0        & -0.0154  & 0.0001   & 0.0174   & 0.0107   & -0.0015  & 0.0037  \\
}
\section{Analysis}
\subsection{Modelling}
\begin{figure}[ht]
\centering
\includegraphics[width=0.95\textwidth]{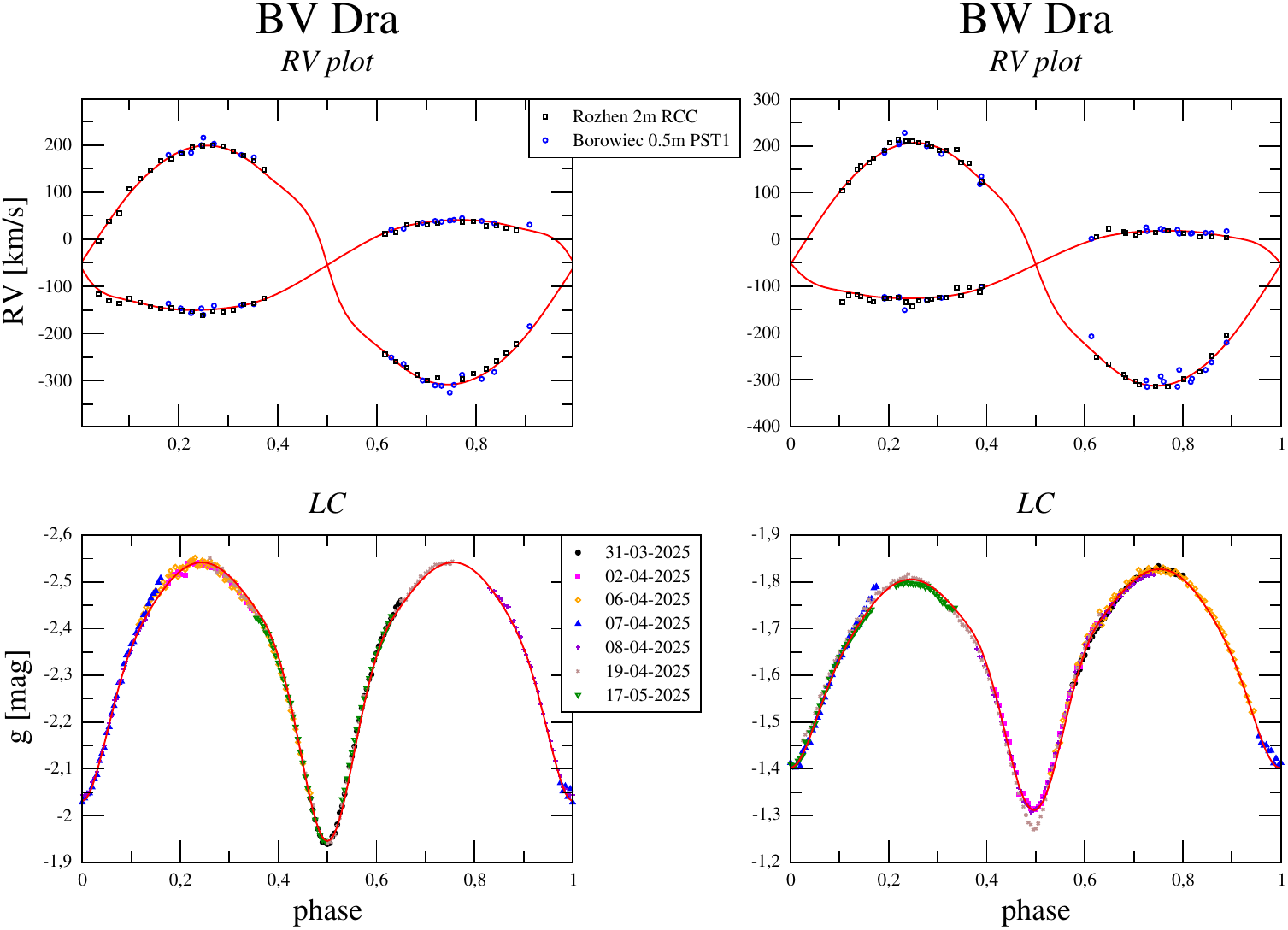}
\caption{Plots show a collection of spectroscopic and photometric data, as well as calculated models obtained using PHOEBE (red lines). Models details are presented in Tab. \ref{Models}, altogether in comparison with results from Kaluzny and Rucinski (1986). 
\textbf{Top row:} Radial velocity plots. Data from Rozhen are marked with black squares, and from PST1 with blue circles.
\textbf{Bottom row:} Light curve plots. Original data was shifted by values from Tab. \ref{shifts_tab}.}
\label{plots}
\end{figure}

Using the data collected, we were able to produce models for both systems. For the modelling process, we used the Wilson-Devinney method (Wilson and Devinney 1971) implemented in the PHOEBE code (Pr\v{s}a and Zwitter 2005). The photometric data were averaged in 300 equally spaced phase bins. The parameters obtained are presented in Tab. \ref{Models}. Radial velocities curves, as well as light curves with fitted models are presented in Fig. \ref{plots}.

Inspecting Tab. \ref{Models} reveals that in the case of some parameters (masses, radii, etc.), we obtain the values that do not fall into the range of the parameter uncertainty from previous studies. The first radial velocities measurements were obtained by Batten and Lu (1986). They used a point source model and fitted sine functions to the data. The same data were analysed by Kalu\.{z}ny and Ruci\'nski (1986) using a full W-D model.
For both eclipsing pairs, radial velocity measurements are difficult due to their short orbital period limiting exposure time and strongly rotationally broadened spectral lines. The masses and radii determined are very sensitive to the amplitude of the RV curves. 
Differences in RV amplitude measurements could result from that in our cross correlation we used wider range of the spectrum wavelength ($\lambda$~4750-6000\r{A}) in comparison with Batten and Lu (1986) measurements ($\lambda$~4000-4280\r{A}). When using the cross-correlation function, a larger spectral range allows for more accurate measurements, as we take into account a significantly larger number of spectral lines. Another difference between the data is the resolution, 12,000 in the case of Batten and Lu (1986) and 35,000 in our case.

In the case of BW Dra, the height difference between two luminosity maxima is easily visible (see Fig. \ref{shifts_pic}). As we rejected the possibility of wrongly shifting the observational data, we looked at another option, the influence of a stellar spot. We found that such a phenomenon as in the LC of BW Dra could be produced by a spot on the primary component with colatitude 28$^\circ$ (measured from the pole), longitude 306$^\circ$, radius 22$^\circ$, and temperature 10\% lower than stellar. 

\MakeTable{lcc|cc}{12.5cm}{Comparison of combined light-curve and spectroscopic solutions for BV Dra and BW Dra. Asterisk indicates fixed value, taken from Kaluzny and Rucinsk (1986). \label{Models}}
{
\hline\noalign{\smallskip}
                   & \textbf{BV Dra}     &                  & \textbf{BW Dra}     &                  \\
\hline\noalign{\smallskip}
Parameter          & Kaluzny and Rucinski& This work        & Kaluzny and Rucinski& This work        \\
                   & (1986)              &                  & (1986)              &                  \\
\hline\noalign{\smallskip}
 $i$ [deg]         & 76.28$\pm$0.32      & 76.17$\pm$0.39   & 74.42$\pm$0.25      & 71.78$\pm$0.13     \\
 $q$               & 0.411$\pm$0.006     & 0.397$\pm$0.006  & 0.280$\pm$0.005     & 0.296$\pm$0.005 \\
 $a$ [R$_\odot$]   & 2.377$\pm$0.016     & 2.583$\pm$0.012  & 1.961$\pm$0.013     & 2.101$\pm$0.011  \\
 $V_\gamma$ [km/s] & -61.2$\pm$1.0       & -54.6$\pm$0.4    & -60.1$\pm$0.4       & -52.7$\pm$1.0   \\
 $K_1$ [km/s]      & 97.2$\pm$1.1        & 95.6             & 71.6$\pm$1.1        & 72.8             \\
 $K_2$ [km/s]      & 236.3$\pm$1.8       & 253.3            & 255.4$\pm$1.9       & 269.6            \\
    &  &  &  & \\
 $T_1$ [K]         & 6245*               & 6245*            & 5980*               & 5980*            \\
 $T_2$ [K]         & 6345$\pm$30         & 6541$\pm$7       & 6164$\pm$10         & 6144$\pm$30      \\
 $\Omega_1$        & 2.345$\pm$0.006     & 2.655$\pm$0.002  & 2.397$\pm$0.006     & 2.427$\pm$0.002  \\
 $L_1/(L_1 + L_2)$ & 0.6734$\pm$0.0015   & 0.6436$\pm$0.0004& 0.7348$\pm$0.0003   & 0.6703$\pm$0.0005\\
    &  &  &  & \\
 $M_1$ [M$_\odot$] & 1.04$\pm$0.02       & 1.35$\pm$0.02    & 0.92$\pm$0.02     & 1.12$\pm$0.03      \\
 $M_2$ [M$_\odot$] & 0.43$\pm$0.01       & 0.55$\pm$0.01    & 0.26$\pm$0.01     & 0.33$\pm$0.01      \\
 $R_1$ [R$_\odot$] & 1.12$\pm$0.01       & 1.21$\pm$0.01    & 0.98$\pm$0.01     & 1.04$\pm$0.01      \\
 $R_2$ [R$_\odot$] & 0.76$\pm$0.01       & 0.81$\pm$0.01    & 0.55$\pm$0.01     & 0.58$\pm$0.01      \\
    &  &  &  & \\
 $n_{obs}$         & 72 (61)             & 77               & 64 (47)           & 108                \\
 method            & c.c.f.              & c.c.f.           & c.c.f.            & c.c.f.             \\
\noalign{\smallskip}\hline
}

\subsection{Light-time effects}
\MakeTable{ccccccc}{12.5cm}{Times of minima for BV Dra obtained from our observations. Epoch 0 is HJD 2444474.3267 (1980-08-22, 19:50), period $p=0.35006764$d. \label{RBT_min}}
{
 \textbf{HJD$_{min}$} & \textbf{UTC$_{min}$}& \textbf{Epoch}   & \textbf{O-C [d]}   & \textbf{P/S} & \textbf{Phot. band} & \textbf{Inst.}\\
\noalign{\smallskip}
\hline\noalign{\smallskip}
 2460400.477416       & 2024-03-30 13:27    & 45494.5 & -0.001832 & S   &  sp & RCC/ESpeRo \\
 2460750.546143       & 2025-03-16 01:06    & 46494.5 & -0.000745 & S   &  sp & PST1  \\
 2460765.772211       & 2025-03-31 06:32    & 46538   & -0.002619 & P   &  g  & RBT   \\
 2460773.648596       & 2025-04-08 03:34    & 46560.5 & -0.002756 & S   &  g  & RBT   \\
 2460774.699640       & 2025-04-09 04:47    & 46563.5 & -0.001915 & S   &  g  & RBT   \\
 2460786.076267       & 2025-04-20 13:50    & 46596   & -0.002487 & P   &  g  & RBT   \\
 2460812.681772       & 2025-05-17 04:22    & 46672   & -0.002122 & P   &  g  & RBT   \\
}
\MakeTable{ccccccc}{12.5cm}{Times of minima for BW Dra obtained from our observations. Epoch 0 is HJD 2440362.875 (1969-05-21, 09:00), period $p=0.292165$d. \label{RBT_minBW}}
{
 \textbf{HJD$_{min}$} & \textbf{UTC$_{min}$}& \textbf{Epoch}   & \textbf{O-C [d]}   & \textbf{P/S} & \textbf{Phot. band} & \textbf{Inst.}\\
\noalign{\smallskip}
\hline\noalign{\smallskip}
 2460401.439225       & 2024-03-31 22:30    & 68586.5 & -0.010548 & S   &  sp & RCC/ESpeRo \\
 2460750.283148       & 2025-03-15 18:45    & 69780.5 & -0.011634 & S   &  sp & PST1  \\
 2460768.539258       & 2025-04-03 00:54    & 69843   & -0.015838 & P   &  g  & RBT   \\
 2460773.650524       & 2025-04-08 03:34    & 69860.5 & -0.017458 & S   &  g  & RBT   \\
 2460774.380936       & 2025-04-08 21:06    & 69863   & -0.017460 & P   &  g  & RBT   \\
 2460785.776840       & 2025-04-20 06:36    & 69902   & -0.015991 & P   &  g  & RBT   \\
 2460813.680353       & 2025-05-18 04:19    & 69997.5 & -0.014235 & S   &  g  & RBT   \\
}
\begin{figure}[ht]
\centering
\includegraphics[width=0.8\textwidth]{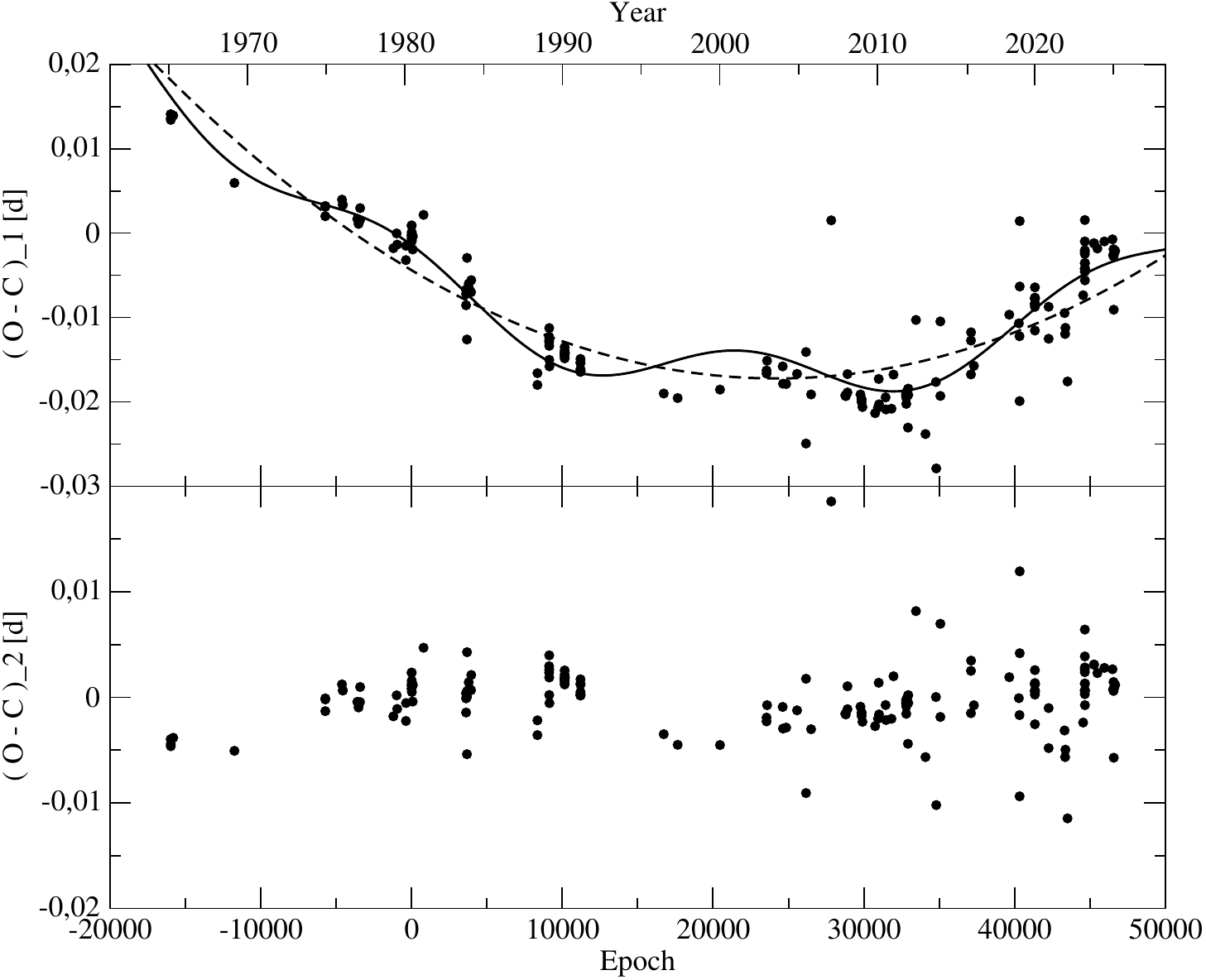}
\caption{\textbf{Upper panel:} (O - C)\_1 curve for BV Dra. The solid and dotted lines represent all contributions and the parabolic part of Equation (1), respectively. \textbf{Lower panel:} Residuals of (O - C)\_2 for the eclipsing binary BV Dra.}
\label{BV_O-C}
\end{figure}
Yang \textit{et al.} (2009) proposed non-linear fitting to the (O-C) curve of BV Dra. This calculation was based on the observations of minima in LC from 1965 to 2005. We tried to repeat that with more data, as we have additional 20 years of observational data. We used times of minima available on the VarAstro database (see also Li\v{s}ka and Skarka, 2014), as well as data from our observations, which are presented in Tab. \ref{RBT_min}. 

Non-linear curve fitting was done using implemented in Gnuplot software method (which uses least squares minimalisation process). We obtain the following equation: 
\begin{equation}
\begin{split}
    (O-C)=-0.0044(\pm0.0004) -10.6(\pm0.5)\times 10^{-7} \times E + 2.19(\pm0.12)\times 10^{-11} \times E^2\\ + 0.0032(\pm0.0004)\times\sin (2.73(\pm0.08)\times10^{-4}\times E +1.88(\pm0.22)),
\end{split}
\end{equation}
where the first part describes quadratic changes (steady increase in period) and the second part describes cyclic changes. The visualisation is presented in Fig. \ref{BV_O-C}.

Our results differ slightly from those obtained by Yang \textit{et al.} (2009): the period increase rate found by us is 0.39 $\pm$ 0.02 s/century (0.25 previously), and the sinusoidal term has a period $P_3=22.1 \pm 0.6 $ yr (instead of $23.6 \pm 0.6$ yr). Following the procedure and the assumptions described by Yang \textit{et al.} (2009) and using the equation from Kwee (1958),
\begin{equation}
    \frac{dP}{P}=3 \left( \frac{M_1}{M_2}-1\right)\frac{dM_1}{M_1},
\end{equation}
we can calculate the mass transfer rate $dm/dt=0.40\times10^{-7} M_{\odot}$yr$^{-1}$. This value is twice as large as the one obtained by Yang \textit{et al.} (2009) $dm/dt=0.20\times10^{-7} M_{\odot}$yr$^{-1}$, as a consequence of the change in period and masses of the components in the BV Dra system.

From the well-known equation describing the mass function, we can obtain the mass and orbital radii of the tertiary body (assuming that the inclination is the same as in the close binary system). Cyclic changes can be explained by a red dwarf with a mass around 0.11 $M_{\odot}$, on an orbit around the BV Dra binary with 9.4 AU. The comparison of the parameters describing changes in the BV Dra period between our and Yang \textit{et al.} (2009) values is presented in Tab. \ref{BV_lite}.

\MakeTable{lcc}{12.5cm}{Parameters for quadratic plus cyclic changes for BV Dra \label{BV_lite}}
{
 \textbf{Parameters} & \textbf{Yang \textit{et al.} (2009)}& \textbf{This work}  \\
\noalign{\smallskip}
\hline\noalign{\smallskip}
$dP/dt$ (d yr$^{-1}$)& $+0.29 \times 10^{-7}$ & $+0.46 (\pm 0.02)\times 10^{-7}$\\
$A$ (d) & $0.0029 (\pm 0.0003)$ & $0.0031(\pm 0.0003)$\\
$P_3$ (yr) & $23.8 (\pm0.6)$ & $22.1 (\pm0.6)$ \\
\hline\noalign{\smallskip}
$dm/dt$ ($M_\odot$ yr$^{-1}$)&  $0.20 \times 10^{-7}$ & $0.39(\pm0.03)\times 10^{-7}$\\
$a_{12}\sin{i'}$ (AU) & $0.5027(\pm 0.0488)$ & $0.522(\pm0.055) $\\
$f(m)$ ($M_\odot$) & $2.24(\pm0.76)\times10^{-4}$ & $2.91(\pm1.07)\times10^{-4}$\\
\hline\noalign{\smallskip}
$M_3$ ($M_\odot$)&  $0.084$ & $0.114\left(^{+0.015}_{-0.019}\right)$ \\
$a_3$ (AU) & $9.1$ & $9.38\left(^{+0.02}_{-0.03}\right)$\\
}

\begin{figure}[ht]
\centering
\includegraphics[width=0.8\textwidth]{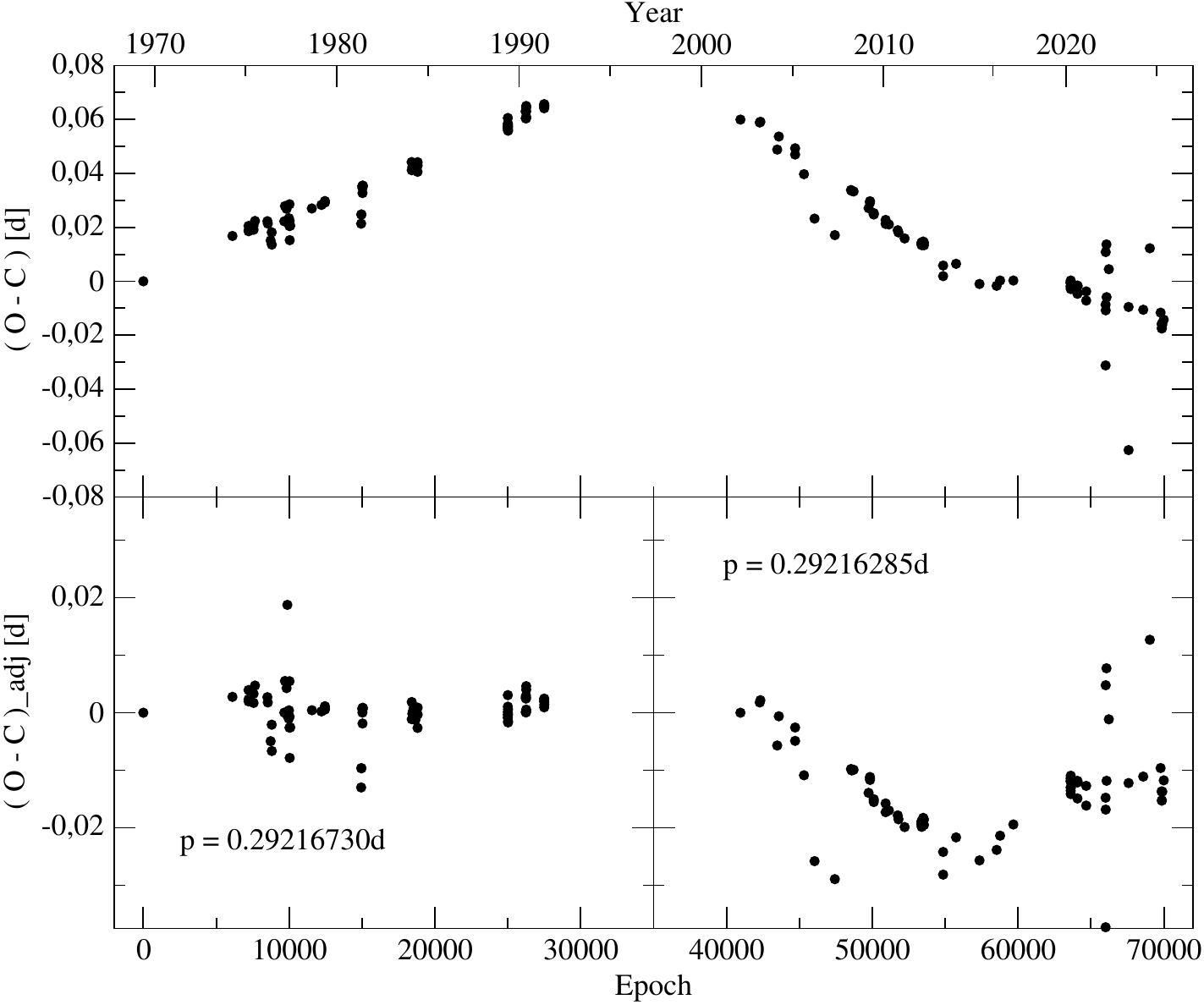}
\caption{\textbf{Upper panel:} (O - C) curve for BW Dra. if single period $p=0.292165$d was used to whole dataset. \textbf{Lower panel:} (O - C) curve adjusted with two different periods: \textbf{left} uses $p=0.29216730$d for epochs from 0 to 30 000 (years 1969-1992), and \textbf{right} uses $p=0.29216285$d for epochs from 40 000 to 70 000 (years 2002-2025)}
\label{BW_O-C}
\end{figure}

Li\v{s}ka and Skarka (2014) point out another interesting phenomenon - antiparallel changes in O-C between BV and BW Dra. Based on data available they proposed an explanation that changes in (O-C) in both systems could be produced by systems' long-period orbital motion. Nevertheless, they pointed out that the probability of such behaviour of the (O-C) curves explained by the maximum radial distance between BV and BW Dra is very low. As we now know, the distance between systems is larger than 1150 AU - the value they used (Batten and Lu, 1986), closer to the value 0.1 pc (Tokarek and Dimitrov, 2025), and thus a longer period could rule out this proposition. 

We propose a different explanation - a break in period. A simple quadratic function cannot be fitted to the (O-C) curve. But, as could be seen in Fig \ref{BW_O-C}, the entire data set can be split into two separate, and by using different periods we obtain a clearer view of the systems behaviour. The difference between these two periods is around $\Delta p=0.4$s. Additionally, in the second part of the data, unusual behaviour can be seen. Combining this with information that we model a spot on BW Dra (thus its an active star system), we can presume that some abrupt event happened between years 1992 and 2002 which caused change in systems behaviour.

\section{Conclusions}
The BV/BW Dra eclipsing binaries are a unique coplanar multiple system. The history of research dates back to 1932. Despite this, our spectroscopic observations are only the second set of data. Forty years after the first observations, we repeated the radial velocity measurements for both pairs. The radial velocity amplitudes obtained for the second component in both pairs are higher than those obtained previously. This is probably due to both the instrument used and the difficulty of observing the object. The instruments differed significantly in resolution and spectral range. The objects, on the other hand, cause problems with their very short orbital periods and broadened spectral lines. Higher RV amplitudes are related to the scale of the systems and result in higher values for some masses and radii.
New eclipse moments of BV Dra confirm and refine Yang's results. Two effects are clearly visible on the O-C diagram: a long-term parabolic change and short-term sinusoidal variability. In case of BW Dra we propose a brake in period as an explanation of the shape of O-C diagram.
\Acknow{J. J. Tokarek gratefully acknowledges observing grant support from the Institute of Astronomy and National Astronomical Observatory, Bulgarian Academy of Sciences. 
This research was funded in part by the National Science Centre, Poland, Grant No. 2021/41/N/ST9/04259. For the purpose of Open Access, the author has applied a CC-BY public copyright licence to any Author Accepted Manuscript (AAM) version arising from this submission. 
NOIRLab IRAF is distributed by the Community Science and Data Center at NSF NOIRLab, which is managed by the Association of Universities for Research in Astronomy (AURA) under a cooperative agreement with the U.S. National Science Foundation. 
The authors are grateful to Tomasz Kwiatkowski, Przemys\l{}aw Bartczak, Aleksander Schwarzenberg-Czerny and our engineer Roman Baranowski, founders of the Pozna\'{n} Spectroscopic Telescope project.}


\begin{references}
\refitem{Aitken, R.G. and Doolittle, E.}{1932}{Carnegie Inst. Washington D.C. Publ. 417}{New General Catalogue of Double Stars within 120{\textdegree} of the North Pole}{0}
\refitem{Batten, A.H., and Hardie, R.H.}{1965}{AJ}{70}{666}
\refitem{Batten, A.H., and Lu, W.}{1986}{PASP}{98}{92}
\refitem{Bailer-Jones, C.A.L., Rybizki, J., Fouesneau, M., \textit{et al.}}{2021}{Astron.J.}{161}{147}
\refitem{Baranowski, R., Smolec, R., Dimitrov, W., \textit{et al.}}{2009}{MNRAS}{396}{2194}
\refitem{Bilir, S., Karata\k{s}, Y., Demircan, O., and Eker, Z.}{2005}{MNRAS}{357}{497-517}
\refitem{Bonev, T., Haralambi, M., Tomov, T., \textit{et al.}}{2017}{BAJ}{26}{67}
\refitem{Currie, M. J., Berry, D. S., Jenness, T., \textit{et al.}}{2014}{ASP Conference Series}{485}{391}
\refitem{Kaluzny, J. and Rucinski, S.M.}{1986}{AJ}{92}{66}
\refitem{Kami\'{n}ski, K., Baranowski, R., Fagas, M., \textit{et al.}}{2014}{Proc. IAU Symp}{301}{437}
\refitem{Kwee, K.K.}{1958}{Bull. Astron. Inst. Neth.}{14}{131}
\refitem{Li\v{s}ka, J., and Skarka, M.}{2014}{OEJV}{169}{38}
\refitem{Poli\'{n}ska, M., Kami\'{n}ski, K., Dimitrov, W., \textit{et al.}}{2019}{Contrib. Astron. Obs. Skaln. Pleso}{49}{484}
\refitem{Pr\v{s}a, A. and Zwitter, T.}{2005}{ApJ}{628}{426}
\refitem{Rovithis, P., and Rovithis-Livaniou, H.}{1983}{Astrophys. Space Sci.}{97}{337}
\refitem{Tody, D.}{1986}{SPIE Conference Series}{627}{733}
\refitem{Tody, D.}{1993}{ASP Conference Series}{52}{173}
\refitem{Tokarek, J.J., and Dimitrov, W.}{2025}{AcA}{75}{62}
\refitem{Wilson, R.E. and Devinney, E.J.}{1971}{ApJ}{166}{605}
\refitem{Yang, Y.-G., L{\"u}, G.-L., Yin, X.-G., \textit{et al.}}{2009}{Astron. J.}{137}{236}
\end{references}
\end{document}